\documentclass[11pt]{article} 
\usepackage{rotating} 
\usepackage{amsbsy} 

\def\ket#1{ $ \left\vert  #1   \right\rangle $ }

\begin{document} 
 
\title{Pure spin--angular momentum coefficients for non--scalar one--particle 
       operators in  
       $jj$--coupling\footnote{Submitted as: New Version Announcement} } 
 
\author{G.\ Gaigalas$^{\,a}$ and S. Fritzsche$^{\,b\,}$\footnote{Email:  
                                 s.fritzsche@physik.uni-kassel.de} \\ 
        \\ 
        $^a$ Institute of Theoretical Physics and Astronomy,    \\ 
        A.\ Go\v{s}tauto 12, Vilnius 2600, Lithuania. 
        \\ 
        $^b$ Fachbereich Physik, Universit\"a{}t Kassel,        \\ 
        Heinrich--Plett--Str. 40, D -- 34132 Kassel, Germany.   \\ 
       \\ 
       } 
 
\maketitle 
 
\date{} 
 
\thispagestyle{empty} 
\enlargethispage{1.0cm} 
\vspace*{-1.0cm} 
 
\begin{abstract}
A \textit{revised} program for generating the spin--angular coefficients in 
relativistic atomic structure calculations is presented. When compared with
our previous version [G.~Gai\-galas, S.~Fritzsche and I.~P.~Grant, 
CPC {\bf 139} (2001) 263], the new version of the \textsc{Anco} program 
now provides these coefficients for both, scalar as well as 
\textit{non--scalar} one--particle operators as they arise frequently in 
the study of transition probabilities, photoionization and electron capture 
processes, the alignment transfer through excited atomic states, collision 
strengths, and in many other investigations. 

The program is based on a recently developed formalism 
[G.~Gaigalas, Z.~Rudzikas, and C.~F.~Fischer, J.\ Phys.\ B \textbf{30} 
(1997) 3747], which combines techniques from second quantization in 
coupled tensorial form, the theory of quasispin, and the use of 
\textit{reduced} coefficients of fractional parentage, in order to derive
the spin--angular coefficients for complex atomic shell structures more
efficiently. By making this approach now available also for non--scalar
interactions, therefore, studies on a whole field of new properties and 
processes are likely to become possible even for atoms and ions with a
complex structure.

 \end{abstract}

\newpage 
 
{\large\bf NEW VERSION SUMMARY} 
 
\bigskip

\textit{Title of program:} ANCO$^{(2)}$ 
 
\bigskip

\textit{Catalogue number:} ADQO
 
\bigskip 
 
\textit{Program Summary URL:} http://cpc.cs.qub.ac.uk/summaries/ADQO

\bigskip  
 
\textit{Program obtainable from:} CPC Program Library,  
     Queen's University of Belfast, N. Ireland.  
 
\bigskip

\textit{Catalogue identifier of previous version:}  ADOO [1]; title:  
                                                    \textsc{Anco}. 
 
\bigskip 
 
\textit{Authors of the original program:} G.~Gaigalas, S.~Fritzsche and  
                                          I.~P.~Grant 
 
\bigskip

\textit{Does the new version supersede the previous one:}  \newline 
     Yes, apart from scalar one-- and two--particle tensor operators, 
     the program now supports also \textit{non--scalar} one--particle
     operators $\hat{A}^k$ for any rank $ k \,>\, 0$. 
      
\bigskip

\textit{Licensing provisions:} None. 
 
\bigskip

\textit{Computer for which the new version has been tested:}   IBM RS 6000, 
     PC Pentium II, AMD Athlon K7.                              \newline 
     {\it Installations:} University of Kassel (Germany).       \newline 
                          Institute of Theoretical Physics and Astronomy 
                          (Lithuania).   \newline 
     {\it Operating systems:} IBM AIX 6.2.2+, Linux 7.1 
      
\bigskip

\textit{Program language used in the new version:} \textsc{Ansi} standard  
     Fortran 90/95. 
 
\bigskip

\textit{Memory required to execute with typical data:} 200 kB. 
 
\bigskip

\textit{No.\ of bits in a word:}  All real variables are parameterized by a 
     \texttt{selected kind parameter}. Currently this parameter is set to  
     double precision (two 32--bit words) for consistency with other  
     components of the \textsc{Ratip} package~\cite{F:2001}. 
 
\bigskip

\textit{Distribution format:} Compressed tar file.

%
%
%
 
\bigskip

 
\bigskip

\textit{Nature of the physical problem:}  \newline 
     The matrix elements of a one--electron tensor operator $\hat{A}^k$  
     of (any) rank $k$ with respect to a set of configuration state functions 
     (CSF) \ket{\gamma_i J_i P_i} can be written as
     $\sum_{ij} \, t^{k}_{ij} (ab)\, (a| \hat{A}^k |b)$ where the  
     \textit{spin--angular coefficients} $t^{k}_{ij}(ab)$ are independent of  
     the operator $\hat{A}^k$, $i,j$ are CSF labels and $a,b$ specify 
     the interacting orbitals. 
     A combination of second--quantization and quasispin methods has been 
     utilized in order to derive and to obtain these angular coefficients 
     for one--electron tensor operator of any rank~\cite{G:1999}. 
     Operators of non--scalar rank, $k\, > \,0$, occur frequently, for instance,
     in the study of transition probabilities, photoionization and electron 
     capture processes, alignment transfer, collision 
     strengths, and elsewhere. 
 
\bigskip

\textit{Reasons for the new version:}  \newline 
     The \textsc{Ratip} package~\cite{F:2001} has been found an efficient 
     tool during recent years, in order to exploit the (relativistic atomic) 
     wave functions from the \textsc{Grasp92} program~\cite{Grasp92} for 
     the computation of atomic properties and processes. 
     Apart from a more efficient set--up of the (Dirac--Coulomb--Breit) 
     Hamiltonian matrix \cite{FFG:2002}, the \textsc{Ratip} program 
     now supports \textit{large--scale} computations of transition  
     probabilities within a \textit{relaxed} orbital basis~\cite{FFD:2000}, 
     Auger rates and angular--anisotropy parameters~\cite{USCKTFK:2001}, and
     of several other quantities. 
     For these properties, the spin--angular coefficients for scalar one-- 
     and two--particle operators are sufficient, as be obtained 
     from the previous version of \textsc{Anco}~\cite{GFG:2001}. However, 
     in order to extent the range of (possible) applications also to other 
     processes such as the atomic photoionization, (radiative) electron 
     capture, or photoexcitation and alignment studies, \textit{non--scalar}
     one--particle operators will occur and need to be treated efficiently. 
     With the presently revised version of \textsc{Anco}, we provide the 
     spin--angular coefficients for such operators, making use of the modern 
     design of the \textsc{Ratip} package in Fortran 90/95~\cite{MR:1996}. 
     Similarly as for all previously implemented components of this package,
     the revised \textsc{Anco} program facilitates the use of large wave 
     function expansions of several ten thousand CSF or even more in the 
     future. 
      
      
\bigskip

\textit{Summary of revisions:}  \newline 
     When compared with the previous CPC version of the \textsc{Anco} 
     program~\cite{GFG:2001}, the following modifications and (new) 
     capabilities have been added:

\begin{enumerate} 
   \item \ 
   The module \texttt{rabs\_recoupling} has been enlarged to include further
   expressions from Ref.\ \cite{GRF:1997}, i.e.\ Eq.\  (14) and (19).
   These expressions are incorporated into the routines 
   \texttt{recoupling\_matrix\_1p\_shell} for calculating the recoupling 
   matrix for the case of CSF with one open shell and into 
   \texttt{recoupling\_matrix\_2p\_shells} for CSF with two open shells,
   respectively. Moreover,
   the subroutine \texttt{recoupling\_C\_3} has been added to derive the 
   $C_3$ coefficients, cf.\ \cite[Eq. (17)]{GRF:1997}.
   
 
   \item \ 
   Several procedures have been added also to the module \texttt{rabs\_anco} 
   to enable the user with a simple and direct access to the spin--angular 
   coefficients. For example, the two routines 
   \texttt{anco\_calculate\_csf\_pair\_1p} and
   \texttt{anco\_calculate\_csf\_matrix\_1p} provide these coefficients
   for any one--particle operator with specified rank, either for a single 
   pair of CSF or for a whole array of such functions, respectively. 
   Both procedures make use of the subroutines 
   \texttt{anco\_one\_particle\_diag} for the diagonal matrix elements and 
   \texttt{anco\_one\_particle\_off} otherwise. In 
   \texttt{anco\_calculate\_csf\_matrix\_1p}, the spin--angular coefficients
   are calculated for any rank $k \,\ge\, 0$ 
   either for the whole matrix or for a sub--matrix with rows from (given)
   \texttt{row\_low} ... \texttt{row\_up} and columns from 
   \texttt{col\_low} ... \texttt{col\_up}. While the procedure
   \texttt{anco\_calculate\_csf\_pair\_1p} returns \texttt{no\_T\_coeff} 
   coefficients directly in the array \texttt{anco\_T\_list}, the coefficients 
   of a whole CSF array are returned in the derived data structure
   \texttt{anco\_pair\_list}; see the header of the module \texttt{rabs\_{}anco} 
   for further details.
   

   \item \ 
   The definition and set--up of properly derived data types 
   in~\cite{GFG:2001} has definitly helped facilitate the data exchange 
   between different components of the \textsc{Ratip} package. 
   These data structures have been used also for extenting the 
   \textsc{Anco} program. For the one--particle coefficients, for example,
   the derived type
   \vspace*{-0.4cm}
   \begin{verbatim} 
   type :: anco_T_coeff 
      integer          :: nu 
      type(nkappa)     :: a, b 
      real(kind=dp)    :: T 
   end type anco_T_coeff 
   \end{verbatim} 
   \vspace*{-0.8cm}
   were introduced already in our previous version, where 
   we used nu = 0 to designate the scalar interaction. The integer \texttt{nu}
   now indicates simply the rank of the (one--particle) tensor. In
   further applications of \textsc{Ratip}, therefore, these coefficients 
   can be easily accessed if the module \texttt{rabs\_{}anco} is 
   properly \texttt{use}d by the additional code.
 
   \item \ 
   A few minor changes are made also in the (interactive) dialog 
   which controls the program as well as for the printout of the 
   spin--angular coefficients. One additional question in the dialog:
   \\[0.2cm]
   \texttt{Generate one-electron angular coefficients for non-scalar 
   interactions ?} 
   \\[0.2cm]
   can first be answered with \texttt{n} or \texttt{N} --- if non--scalar 
   coefficients are not requested. If answered by \texttt{y} or \texttt{Y}, 
   the additional question:
   \\[0.2cm]
   \texttt{Generate GRASP92-like d coefficients for non-scalar interactions ?} 
   \\[0.2cm]
   arise and help the user to select \textsc{Grasp92}--like $d^{\,k}_{ab}(rs)$ 
   coefficients, such as previously obtained from the \textsc{Mct} component
   of \textsc{Grasp}, or 'pure' angular coefficients (as utilized within 
   \textsc{Ratip}). Finally, the prompt
   \\[0.2cm]
   \texttt{Enter the rank of the tensor:} 
   \\[0.2cm]
   requests the user to specify the rank of the one--particle operator. 
   
   \item \ 
   As previously, the \textsc{Anco} program generates two output files;
   while the \texttt{.sum} \textsc{Anco} summary file provides a short 
   summary of the computations, the spin--angular coefficients and all 
   necessary quantum numbers for their classification are stored in the 
   \texttt{.vnu} file. The format for printing the $d^{\,k}_{rs}(ab)$ and 
   $t^{\,k}_{rs}(ab)$ is very similar to each other, apart from the sorting
   process which is used in \textsc{Grasp92}~\cite{Grasp92} and which is not
   done by \textsc{Anco}. 
\end{enumerate} 

     As before, the source code of the \textsc{Anco} component is distributed  
     together with the source of all other components of \textsc{Ratip}
     in order to facilitate the installation and to save the user from
     additional adaptions to be made. The whole package is provided as a 
     compressed \texttt{Ratip\_anco.tar.Z} 
     archive file. On a \textsc{Unix} (or any compatible) system, the two 
     commands \texttt{uncompress Ratip\_anco.tar.Z} and
     \texttt{tar xvf Ratip\_anco.tar} will reconstruct the file structure.
     The obtained \texttt{ratip} root directory then obtains the source code, 
     the file \texttt{make-anco} for generating the executable \texttt{xanco}
     and the subdirectory \texttt{test-anco} for the test example. Details 
     of the installation are explained also in the \texttt{Read.me} file 
     in the \texttt{ratip} root directory 
     
\bigskip

\textit{Restrictions onto the complexity of the problem:}  \newline 
     For all subshells with $ j \ge 11/2$ (i.e.\ $h_{11/2}-, \, i-, \, ...$ 
     electrons), the maximal number of \textit{equivalent electrons} is 
     restricted to two.  
        
\bigskip

\textit{Typical running time:}  \newline 
     This strongly depends on the system and the size of the wave  
     function expansion to be considered. Our test case, which is distributed
     with the code in the subdirectory \texttt{test-anco}, required
     32 seconds on a 1400 MHz AMD Athlon K7/1400T. Typically, \textsc{Anco}
     calculates about 10,000 angular coefficients per second.
     
\bigskip

\textit{Unusual features of the program:} \newline 
     \textsc{Anco} has been designed as component of the \textsc{Ratip} 
     package~\cite{F:2001} for calculating a variety of 
     \textit{relativistic atomic transition and ionization properties}. 
     Owing to the careful use of \texttt{allocatable} and \texttt{pointer} 
     arrays, there is (almost) no restriction on the size or any dimension 
     of the ''physical problem'' apart from the computer ressources themselves.

\bigskip 

\end{document}